\documentclass[12pt]{article}

\usepackage{amsthm,amsmath,mathrsfs,amsfonts,amssymb,mathtools,nicefrac}
\usepackage{graphicx,caption,subcaption}
\usepackage{enumerate}
\usepackage{natbib} \RequirePackage[colorlinks,citecolor=blue,urlcolor=blue]{hyperref}
\usepackage{url} 

\usepackage[multiple]{footmisc} \usepackage[capposition=top]{floatrow}

\usepackage{adjustbox}
\usepackage{multirow}

\usepackage{amsmath}
\usepackage{accents}
\newlength{\dhatheight}

\newlength{\dcheckheight}
\newcommand{\doublecheck}[1]{%
    \settoheight{\dcheckheight}{\ensuremath{\check{#1}}}%
    \addtolength{\dcheckheight}{-0.22ex}%
    \check{\vphantom{\rule{1pt}{\dcheckheight}}%
    \smash{\check{#1}}}}

\newcommand{\blind}{1}
\newtheorem{assumption}{Assumption}

\newtheorem{theorem}{Theorem}

\theoremstyle{definition}

\def\N{n}
\def\T{m}

\begin{document}

\def\spacingset#1{\renewcommand{\baselinestretch}{#1}\small\normalsize} \spacingset{1}

\if1\blind
{
  \title{\bf {Bootstrap inference for fixed-effect models}}
  \author{ Ayden Higgins\thanks{Address: University of Cambridge, Faculty of Economics, Austin Robinson Building, Sidgwick Avenue, Cambridge CB3 9DD, United Kingdom. E-mail: \texttt{amh239@cam.ac.uk}.}\\ {\small Faculty of Economics} \\ {\small University of Cambridge } \and Koen Jochmans\thanks{Address: Toulouse School of Economics, 1 esplanade de l'Universit\'e, 31080 Toulouse, France. E-mail: \texttt{koen.jochmans@tse-fr.eu}. \newline 
 Support from the European Research Council through grant n\textsuperscript{o} 715787 (MiMo), and from the French Government and the ANR under the Investissements d' Avenir program, grant ANR-17-EURE-0010 is gratefully acknowledged.} \\  {\small Toulouse School of Economics} \\ {\small University of Toulouse Capitole}
    }
\date{\small \today}
  \maketitle
} \fi

\if0\blind
{
  \bigskip
  \bigskip
  \bigskip
  \begin{center}
    {\bf {\normalsize BOOTSTRAP INFERENCE IN FIXED-EFFECT MODELS}}
\end{center}
  \medskip
} \fi

\vspace{-.5cm}
\begin{abstract}
\noindent
The maximum-likelihood estimator of nonlinear panel data models with fixed effects is consistent but asymptotically-biased under rectangular-array asymptotics. The literature has thus far concentrated its effort on devising methods to correct the maximum-likelihood estimator for its bias as a means to salvage standard inferential procedures. Instead, we show that the parametric bootstrap replicates the distribution of the (uncorrected) maximum-likelihood estimator in large samples. This justifies the use of confidence sets constructed via standard bootstrap percentile methods. No adjustment for the presence of bias needs to be made.

\end{abstract}

\noindent
{\bf JEL Classification:}  
C23	 

\medskip
\noindent
{\bf Keywords:}  
bootstrap,
fixed effects,
incidental parameter problem,
inference, 
panel data


\spacingset{1.45}

\renewcommand{\theequation}{\arabic{section}.\arabic{equation}}  \setcounter{equation}{0}

\section*{Introduction}

The maximum-likelihood estimator of models for panel data is well known to perform poorly when fixed effects are included. The estimator is inconsistent under asymptotics where the number of individuals, $\N$, grows large while the number of time periods, $\T$, is held fixed (\citealt{NeymanScott1948}). In fact, many parameters of interest are simply not (point) identified in such a setting (see, e.g., \citealt{HonoreTamer2006}). Maximum likelihood is, however, consistent under so-called rectangular-array asymptotics, where $\N$ and $\T$ grow large at the same rate \citep{LiLindsayWaterman2003}. Nevertheless, it is asymptotically biased, in general. This implies that confidence sets based on a naive normal approximation to the distribution of the maximum-likelihood estimator have incorrect coverage, even in large samples. 

Over the last two decades substantial effort has been devoted to devising procedures that remove the asymptotic bias, thereby recentering the limit distribution around zero and restoring the validity of conventional inference procedures based on it. A discussion of this literature as well as an overview of many available approaches is given by \cite{ArellanoHahn2007}.\footnote{Approaches to correct the maximum-likelihood estimator, either via analytical formulae or a jackknife, are considered by \cite{HahnNewey2004}, \cite{HahnKuersteiner2011}, and \cite{DhaeneJochmans2015}. Adjustments to the (profile) likelihood or score equation have been considered by \cite{HahnNewey2004} and  \cite{ArellanoHahn2006}. Strategies based on simulation are discussed in \cite{DhaeneJochmans2015b} and  \cite{KimSun2016}. 
} 
Theoretical guidelines on which bias-correction method to use and on how to select their respective tuning parameters are mostly absent. This is inconvenient because, although all proposals lead to estimators with the same (first-order) asymptotic properties, they vary greatly in ease of implementation and their effectiveness at removing bias in finite samples can be quite heterogeneous. 

The current paper shows that, under rectangular-array asymptotics, the parametric bootstrap consistently estimates the distribution of the (uncorrected) maximum-likelihood estimator, including its asymptotic bias. This implies that confidence sets constructed using the percentile method have correct coverage in large samples. Thus, bias correction is not needed. 
The same conclusion is true for averages over the fixed effects, such as their moments or average marginal effects (\citealt{Chamberlain1984}).

In its simplest form, inference based on the bootstrap only requires a routine to compute the maximum-likelihood estimator.\footnote{Corrections to the estimator require first estimating the asymptotic bias. The latter depends on moments and cross-moments of higher-order derivatives of the likelihood, which can be cumbersome to derive and compute. Adjustments to the (profile) likelihood have the additional inconvenience that they can be difficult to maximize whereas modified (profile) score equations may have no or multiple roots. An example where this problem arises is discussed in \cite{DhaeneJochmans2016}.} It is useful to stress that, in spite of the presence of possibly many fixed effects, conventional numerical optimization is, in fact, straightforward, by exploiting the sparsity of the Hessian matrix.\footnote{The usefulness of partitioned-inverse formulae in models with many parameters has been mentioned before; \cite{PrenticeGloeckler1978} and \cite{Chamberlain1980} did so in the context of duration models and binary-choice models, respectively. It is not clear that it is widely appreciated, however, as estimation with fixed effects is often said to be computationally demanding or even judged to be infeasible; see, e.g., the discussion on computation in \cite{KimSun2016}.} Furthermore, because many popular fixed-effect specifications such as probit and tobit models involve likelihood functions that are globally concave, finding the global maximizer requires only a few iterations. Finally, an excellent starting value for the bootstrap maximum-likelihood estimator comes in the form of the maximum-likelihood estimator based on the original data, as the latter is used to generate the  bootstrap samples.


In Section 1 we present the setting and state our objectives. In Section 2 we describe our bootstrap procedures. In Section 3 we investigate the performance of the bootstrap in three examples through theoretical calculations and simulations. In Section 4 we discuss numerical computation via an efficient Newton-Raphson routine. In Section 5 we collect all the assumptions and formal results that underlie our claims about the validity of the bootstrap. Concluding remarks end the paper. An appendix contains proofs. Additional technical results are collected in a supplement.

\section{Maximum-likelihood estimation}
Suppose that we have data on $\N$ independent stratified observations $\lbrace y_i,y_{i-},x_i\rbrace$, with $y_i\coloneqq(y_{i1},\ldots,y_{i\T})$, $y_{i-}=(y_{i(1-p)},\ldots, y_{i0})$, and $x_i\coloneqq(x_{i1},\ldots,x_{i\T})$. We consider models where the conditional density of $y_i$ given $y_{i-}$ and $x_i$ (relative to some dominating measure) is given by
$$
\prod_{t=1}^\T
f(y_{it} \vert y_{it-1},\ldots, y_{it-p},x_{it}; \varphi_0,\eta_{i0}),
$$
and $f$ is known up to the finite-dimensional parameters $\varphi_0$ and $\eta_{i0}$. This framework covers autoregressive processes (of order $p$), for which $y_{i-}$ serves as the initial condition, as well as models with  exogenous covariates, $x_i$. In what follows we will treat both the initial condition and the covariates as fixed.

It is convenient to introduce the shorthand 
$$
\ell(\varphi,\eta_i \vert z_{it})\coloneqq
\log 
f(y_{it} \vert y_{it-1},\ldots, y_{it-p},x_{it}; \varphi,\eta_i),
$$
where $z_{it}\coloneqq(y_{it},y_{it-1},\ldots, y_{it-p},x_{it})$. The maximum-likelihood estimator is
$$
(\hat{\varphi},\hat{\eta}_1,\ldots,\hat{\eta}_\N) \coloneqq
\underset{\varphi,\eta_1,\ldots,\eta_\N}{\arg\max} 
\sum_{i=1}^\N \sum_{t=1}^\T \ell(\varphi,\eta_i \vert z_{it}).
$$
In sufficiently regular models we have, as $\N,\T\rightarrow\infty$ with $\nicefrac{\N}{\T}\rightarrow \gamma^2$ for some finite $\gamma$, that
\begin{equation} \label{eq:mlelimit}
\sqrt{\N\T}(\hat{\varphi}-\varphi_0)
\overset{L}{\rightarrow} {N}(\gamma \beta,\varSigma),
\end{equation}
for a (non-random) bias term $\beta$ and 
$$
\varSigma\coloneqq
-
\lim_{\N,\T\rightarrow\infty}
\frac{1}{\N\T} \sum_{i=1}^\N \sum_{t=1}^\T
\mathbb{E}
\left( 
\frac{\partial^2 \ell (\varphi_0,\eta_{i0}\vert z_{it})}{\partial\varphi\partial\varphi^\prime}
-
\rho_i 
\frac{\partial^2 \ell (\varphi_0,\eta_{i0}\vert z_{it})}{\partial\eta_i\partial\varphi^\prime}
\right)
^{-1},
$$
with
$$
\rho_i
 \, \coloneqq \, 
\left(
\lim_{\T\rightarrow\infty} \frac{1}{\T} \sum_{t=1}^\T
\mathbb{E}
\left(\frac{\partial^2 \ell (\varphi_0,\eta_{i0}\vert z_{it})}{\partial\varphi\partial\eta_i^\prime}\right) 
\right)
\left(
\lim_{\T\rightarrow\infty} \frac{1}{\T} \sum_{t=1}^\T
\mathbb{E}
\left(\frac{\partial^2 \ell (\varphi_0,\eta_{i0}\vert z_{it})}{\partial\eta_i\partial\eta_i^\prime}\right)
\right)
^{-1},
$$
is the inverse of the Fisher information for $\varphi$; see \cite{HahnNewey2004} and \cite{HahnKuersteiner2011}. 

An implication of \eqref{eq:mlelimit} is that confidence regions based on the limit distribution have to account for the bias term $\beta$ in order to have correct coverage unless $\nicefrac{\N}{\T}$ is close to zero, which is not the case in most applications.
Corrections to the estimator have the generic form 
$$
\hat{\varphi} - \frac{\hat{\beta}}{\T},
$$
where $\hat{\beta}$ is an estimator of $\beta$.
Such corrections recenter the estimator's limit distribution around zero, thereby restoring the validity of conventional inference procedures based on it.

We may also be interested in parameters of the form
$$
\Delta \coloneqq
\lim_{\N,\T\rightarrow\infty}
\frac{1}{\N\T} \sum_{i=1}^\N \sum_{t=1}^\T
\mathbb{E}(\mu(z_{it},\varphi_0,\eta_{i0})),
$$
for a chosen function $\mu$. Average marginal effects (as discussed in \citealt{Chamberlain1984}) or moments of the fixed effects are typical examples. The maximum-likelihood estimator of $\Delta$ is 
$$
\hat{\Delta} \coloneqq
\frac{1}{\N\T} \sum_{i=1}^\N \sum_{t=1}^\T \mu(z_{it},\hat{\varphi},\hat{\eta}_i)
$$
which, similar to $\hat{\varphi}$, also suffers from asymptotic bias. Moreover,
$$
\sqrt{\N\T}(\hat{\Delta}-\Delta)
\overset{L}{\rightarrow} N(\gamma\nabla,\sigma^2).
$$
The form of the bias, $\nabla$, is complicated. Expressions for it (and estimators of it) can be found in \cite{HahnNewey2004} and \cite{DhaeneJochmans2015}.  The asymptotic variance is
$$
\sigma^2\coloneqq
\lim_{\N,\T\rightarrow\infty} \frac{1}{\N\T}
\sum_{i=1}^\N \sum_{t=1}^\T
\mathbb{E}
\left(
{\textstyle \sum_{j=-\infty}^{+\infty}}
\upsilon_{it}^{\vphantom{2}} \, \upsilon_{it-j}^{\vphantom{2}}
+
\omega_{it}^2
\right).
$$
Here the term involving $\upsilon_{it}\coloneqq \mu(z_{it},\varphi_0,\eta_{i0}) - \mathbb{E}(\mu(z_{it},\varphi_0,\eta_{i0}))$ is the long-run variance of the infeasible estimator that presumes the parameters to be known. The second term is the variance of  
$$
\omega_{it}\coloneqq
\varpi^\prime \varSigma
\left(
\frac{\ell(\varphi_0,\eta_{i0} \vert z_{it})}{\partial \varphi}
-\rho_i \frac{\ell(\varphi_0,\eta_{i0} \vert z_{it})}{\partial \eta}
\right)
-
\varrho_i
\frac{\partial \ell(\varphi_0,\eta_{i0} \vert z_{it})}{\partial \eta_i},
$$
where
$$
\varpi
\coloneqq
\lim_{\N,\T\rightarrow\infty} \frac{1}{\N\T}
\sum_{i=1}^\N \sum_{t=1}^\T
\mathbb{E}
\left( 
\frac{\partial \mu(z_{it},\varphi_0,\eta_{i0})}{\partial \varphi}
-
\rho_i
\frac{\partial \mu(z_{it},\varphi_0,\eta_{i0})}{\partial \eta}
\right)
$$
and
$$
\varrho_i
\coloneqq
\left(
\lim_{\T\rightarrow\infty}
\frac{1}{\T} \sum_{t=1}^\T
\mathbb{E}
\left(
\frac{\partial \mu(z_{it},\varphi_0,\eta_{i0})}{\partial \eta_i^\prime} 
\right)
\right)
\left(
\lim_{\T\rightarrow\infty}
\frac{1}{\T} \sum_{t=1}^\T
\mathbb{E}
\left(
\frac{\partial^2 \ell(\varphi_0,\eta_{i0} \vert z_{it})}{\partial \eta_i \partial \eta_i^\prime}
\right) 
\right)^{-1}.
$$
The term making up the second contribution to $\sigma^2$ reflects the fact that the parameters of the model need to be estimated in a first step to be able to estimate $\Delta$.


\section{Bootstrap inference}

The (parametric) bootstrap we consider imposes the data generating process implied by the maximum-likelihood estimator. A bootstrap observation $y_{i}^* \coloneqq(y_{i1}^*,\ldots, y_{i\T}^*)$ can be generated recursively by drawing $y_{it}^*$ from the fitted transition density obtained from the original data, i.e.,  
$$
f( y_{it}^* \vert y_{it-1}^*,\ldots, y_{it-p}^*,x_{it};\hat{\varphi},\hat{\eta}_i).
$$
The initial condition, like the covariates, is held fixed, i.e.,  $y_{i-}^* = y_{i-}^{\vphantom{*}}$. The associated maximum-likelihood estimator  is
$$
(\hat{\varphi}^*,\hat{\eta}_1^*,\ldots,\hat{\eta}_\N^*) \coloneqq
\underset{\varphi,\eta_1,\ldots,\eta_\N}{\arg\max} 
\sum_{i=1}^\N \sum_{t=1}^\T \ell(\varphi,\eta_i \vert z_{it}^*),
$$
with $z_{it}^*\coloneqq(y_{it}^*,y_{it-1}^*,\ldots, y_{it-p}^*,x_{it})$.

The main observation of this paper is that, in regular situations, 
\begin{equation} \label{eq:bstraplimit}
\sqrt{\N\T}(\hat{\varphi}^*-\hat{\varphi}) 
\overset{L^*}{\rightarrow} {N}(\gamma\beta,\varSigma),
\end{equation} 
as $\N,\T\rightarrow\infty$ with $\nicefrac{\N}{\T}\rightarrow \gamma^2$. Throughout, we use $\overset{L^*}{\rightarrow}$ to denote weak convergence of the bootstrap measure. 
Equations \eqref{eq:mlelimit} and \eqref{eq:bstraplimit} reveal that the bootstrap distribution is consistent for the distribution of the maximum-likelihood estimator. Importantly, the bootstrap mimics the asymptotic bias. 

It follows from \eqref{eq:bstraplimit} that asymptotically-valid confidence intervals can be constructed by the usual percentile method, without the need to bias-correct the maximum-likelihood estimator (or, indeed, its bootstrap counterpart). For example, for a chosen vector of conformable dimension $c$,
$$
\left\lbrace 
c^\prime\varphi : 
c^\prime(\hat{\varphi} - \varphi) 
\leq 
 q^*_{1-\alpha}
\right\rbrace
$$
is an upper one-sided confidence interval for the linear combination $c^\prime \varphi_0$ with confidence level $(1-\alpha)$ (in large samples) when setting
$$
q_\alpha^* =
\inf
\left\lbrace q^*:
\alpha \leq \mathbb{P}^*(c^\prime (\hat{\varphi}^* -  \hat{\varphi}) \leq q^*)
\right\rbrace.
$$
The notation $\mathbb{P}^*$ refers to a probability computed with respect to the bootstrap measure, i.e, conditional on the sample. Thus, the critical value $q^*_\alpha$ is the $\alpha$-th quantile of the bootstrap distribution of $c^\prime (\hat{\varphi}^* -  \hat{\varphi})$. 
A two-sided (equal-tailed) confidence interval with the same level of confidence is given by
$$
\left\lbrace 
c^\prime\varphi : 
c^\prime\hat{\varphi} - q^*_{1-\nicefrac{\alpha}{2}}
\leq 
c^\prime\varphi 
\leq
c^\prime\hat{\varphi} - q^*_{\nicefrac{\alpha}{2}}
\right\rbrace.
$$
In both cases, construction of the confidence interval only requires a routine to calculate the maximum-likelihood estimator. 

The conditions underlying \eqref{eq:mlelimit} and \eqref{eq:bstraplimit} imply the consistency of the plug-in estimator 
$\hat{\varSigma}$ 
and of its bootstrap counterpart $\hat{\varSigma}^*$
for the inverse Fisher information $\varSigma$. Consequently, we may equally rely on the percentile $t$-method to perform inference. For example, the set
$$
\left\lbrace 
c^\prime\varphi : 
(c^\prime \, \hat{\varSigma} \, c)^{-\nicefrac{1}{2}}c^\prime (\hat{\varphi}^* -  \hat{\varphi})
\leq 
q^*_{1-\alpha} 
\right\rbrace,
$$
now with 
$$
q_\alpha^* =
\inf
\left\lbrace q^*:
\alpha \leq \mathbb{P}^* \big( (c^\prime \, \hat{\varSigma}^* \, c)^{-\nicefrac{1}{2}}c^\prime (\hat{\varphi}^* -  \hat{\varphi}) \leq q^*\big)
\right\rbrace,
$$
is an upper one-sided confidence set for $c^\prime\varphi_0$ with confidence level $(1-\alpha)$. Its construction is based on the familiar studentized $t$-statistic. For multivariate linear combinations $C^\prime \varphi_0$, where $C$ is a conformable matrix, the set
$$
\left\lbrace 
C^\prime\varphi : 
(\hat{\varphi}^* -  \hat{\varphi})^\prime C \, (C^\prime  \hat{\varSigma} \, C)^{-1}C^\prime (\hat{\varphi}^* -  \hat{\varphi})
\leq 
q^*_{1-\alpha} 
\right\rbrace,
$$
is based on a quadratic form and, hence, ellipsoidal in shape. Here to ensure coverage of ($1-\alpha$) in large samples we use
$$
q_\alpha^* =
\inf
\left\lbrace q^*:
\alpha \leq \mathbb{P}^*\big((\hat{\varphi}^* -  \hat{\varphi})^\prime C \, (C^\prime \, \hat{\varSigma}^* \, C)^{-1}C^\prime (\hat{\varphi}^* -  \hat{\varphi}) \leq q^*\big)
\right\rbrace,
$$
which is the $\alpha$-th quantile of the distribution of the bootstrap version of the quadratic form on which the confidence set is based.

Inference on $\Delta$ may equally be done via the bootstrap. Given a bootstrap sample and the associated maximum-likelihood estimator, we construct the corresponding plug-in estimator
$$
\hat{\Delta}^* \coloneqq
\frac{1}{\N\T} \sum_{i=1}^\N \sum_{t=1}^\T \mu(z_{it}^*,\hat{\varphi}^*,\hat{\eta}^*_i).
$$
The bootstrap distribution of $\sqrt{\N\T}(\hat{\Delta}^*-\hat{\Delta})$ mimics the distribution of $\sqrt{\N\T}(\hat{\Delta}-\Delta)$, in large samples, i.e.,
\begin{equation*} 
\sqrt{\N\T}(\hat{\Delta}^*-\hat{\Delta}) 
\overset{L^*}{\rightarrow} {N}(\gamma\nabla,\sigma^2),
\end{equation*} 
as $\N,\T\rightarrow\infty$ with $\nicefrac{n}{m}\rightarrow \gamma^2$. The construction of confidence intervals for $\Delta$ is then completely analogous to before. 


\section{Examples}

\paragraph{Many normal means}
In the classic problem of \cite{NeymanScott1948} we observe independent variables
$$
z_{it} \sim N(\eta_{i0},\varphi_0).
$$
Maximum likelihood estimates the mean parameters by the within-strata sample averages $\overline{z}_i\coloneqq\nicefrac{1}{\T} \sum_{t=1}^\T z_{it}$ and the common variance parameter by 
$$
\hat{\varphi} = \frac{1}{\N\T} \sum_{i=1}^\N \sum_{t=1}^\T (z_{it}-\overline{z}_i)^2.
$$
It is well-known that, in this case, 
$$
\sqrt{\N\T}(\hat{\varphi}-\varphi_0) \overset{L}{\rightarrow} N(-\gamma \varphi_0, 2\varphi_0^2),
$$
under rectangular-array asymptotics. Starting from the fact that 
$\N\T \, {\hat{\varphi}} / {\varphi_0}\sim \chi^2_{\N(\T-1)}$ the exact distribution of the maximum-likelihood estimator can be derived. We find that 
$$
\sqrt{\N\T}(\hat{\varphi}-\varphi_0)
\sim
\mathrm{Gamma}\left(-\sqrt{\N\T}\varphi_0 ,\frac{\N(\T-1)}{2}, \frac{2\varphi_0}{\sqrt{\N\T}} \right),
$$
where $\mathrm{Gamma}(\vartheta_1,\vartheta_2,\vartheta_3)$ refers to the Gamma distribution with location $\vartheta_1$, shape $\vartheta_2$ and scale $\vartheta_3$. It is readily verified that the mean and variance of this distribution are equal to
$$
-\sqrt{\frac{\N}{\T}} \varphi_0, \qquad 2\varphi_0^2 \left(1 - \frac{1}{\T} \right),
$$
respectively. 

In this example, the bootstrap independently samples $z_{it}^*\sim N(\overline{z}_i,\hat{\varphi})$. The associated maximum-likelihood estimators are $\overline{z}_i^*$ and
$$
\hat{\varphi}^* 
= 
\frac{1}{\N\T} \sum_{i=1}^\N \sum_{t=1}^\T (z_{it}^* - \overline{z}_i^*)^2.
$$
Conditional on the data, the latter estimator follows the same Gamma distribution as above, only with $\varphi_0$ replaced by $\hat{\varphi}$. Noting that we can write
$
\sqrt{\N\T}(\hat{\varphi}  - \varphi_0) = - \sqrt{\nicefrac{\N}{\T}}\, \varphi_0 + \epsilon,
$
for a mean-zero random variable $\epsilon=O_P(1)$, this implies that
$$
\sqrt{\N\T}(\hat{\varphi}^*-\hat{\varphi})
\sim
\mathrm{Gamma}\left(-\left(\sqrt{\N\T}\varphi_0 - \sqrt{\frac{\N}{\T}} \varphi_0 + \epsilon \right),\frac{\N(\T-1)}{2}, \frac{2\varphi_0}{\sqrt{\N\T}}\left(1-\frac{1}{\T} \right) + \frac{2\epsilon}{\N\T} \right)
$$
conditional on the sample. Its mean and variance are
$$
-\sqrt{\frac{\N}{\T}}\varphi_0  + \frac{1}{\T} \left( \sqrt{\frac{\N}{\T}} \varphi_0  - \epsilon \right),
\qquad
2\varphi_0^2 \left(1-\frac{2}{\T} + \frac{1}{\T^2} \right)  
+
O_P\left(\frac{1}{\T}\right),
$$
which, to first order, agree with the corresponding moments of the maximum-likelihood estimator.

The studentized maximum-likelihood estimator follows a (translated) inverse-Gamma distribution, mirrored about the origin. Moreover,  
$$
-
\sqrt{\N\T} \, \frac{(\hat{\varphi}-\varphi_0)}{\sqrt{2\hat{\varphi}^2}}
\sim
\text{Inverse-Gamma}
\left(-\sqrt{\frac{\N\T}{2}}, \frac{\N(\T-1)}{2},  \sqrt{\frac{\N\T}{2}} \, \frac{\N\T}{2}\right).
$$
This distribution is pivotal, and the bootstrap replicates it exactly. Thus, at least in this example, the percentile-$t$ method yields confidence intervals whose probability of covering $\varphi_0$ can be controlled exactly.


A first-order correction to $\hat{\varphi}$ based on a plug-in estimator of its asymptotic bias is
$$
\check{\varphi}\coloneqq \hat{\varphi} + \frac{\hat{\varphi}}{\T}.
$$
It is interesting to compare the performance of confidence intervals for $\varphi_0$ based on bias correction with those obtained via the bootstrap. The bias-correction approach uses the large-sample approximation 
$$
\sqrt{\N\T}\, \frac{(\check{\varphi}-\varphi_0)}{\sqrt{2\hat{\varphi}^2}}
\overset{L}{\rightarrow} N(0,1).
$$
Its coverage accuracy can be evaluated for any given sample size from the observation that
$$
-\sqrt{\N\T} \, \frac{(\check{\varphi}-\varphi_0)}{\sqrt{2\hat{\varphi}^2}}
\sim
\text{Inverse-Gamma}
\left(-\sqrt{\frac{\N\T}{2}} \left(1+\frac{1}{\T} \right), \frac{\N(\T-1)}{2},  \sqrt{\frac{\N\T}{2}} \, \frac{\N\T}{2}\right).
$$
Notice that this distribution coincides with that of the studentized maximum-likelihood estimator up to the location parameter; the current distribution being located closer to zero. An alternative in this particular example is to studentize the bias-corrected estimator using $\sqrt{2\check{\varphi}^2}$. We find that
$$
-\sqrt{\N\T} \, \frac{(\check{\varphi}-\varphi_0)}{\sqrt{2\check{\varphi}^2}}
\sim
\text{Inverse-Gamma}
\left(-\sqrt{\frac{\N\T}{2}} , \frac{\N(\T-1)}{2},  \sqrt{\frac{\N\T}{2}} \, \frac{\N\T}{2} \left(\frac{\T}{\T+1}\right)\right).
$$
Here, there is no change in the location parameter (compared to maximum likelihood) but, rather, in the scale parameter. This, then, affects the  entire shape of the sampling distribution.

\begin{figure}[h!]
\centering 
\caption{Many normal means: Sampling densities and distributions}
\includegraphics[width=.95\textwidth]{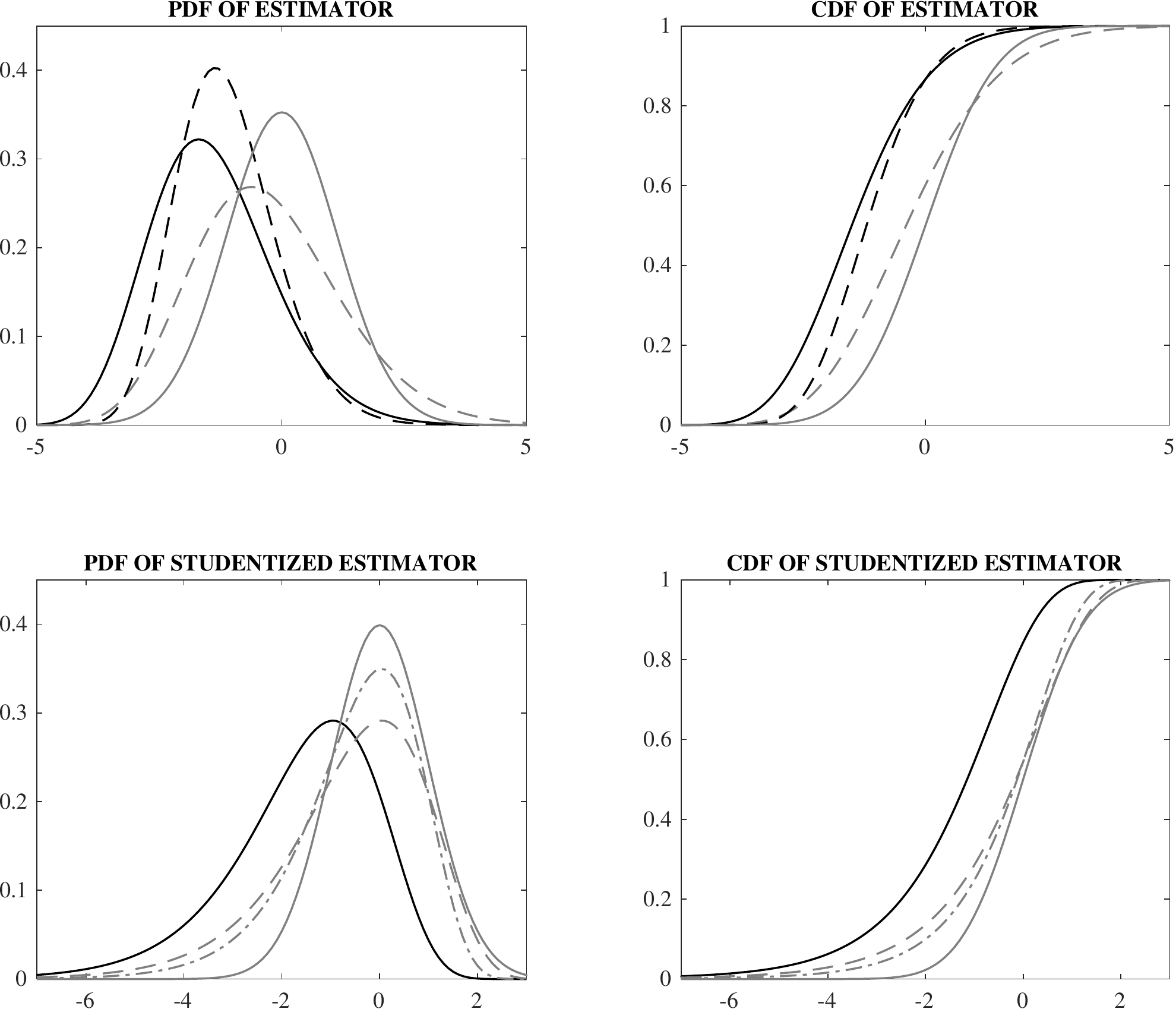}
\vspace{-0.25cm} 
\floatfoot{Upper panel: Density functions (left plot) and cumulative distributions (right plot) of $\hat{e}$ (solid black curve), $\hat{e}^*$ (dashed black curve), and $\check{e}$ (dashed grey curve), together with the normal density with zero mean and variance $2\varphi_0$ (solid grey curve). Lower panel: Density functions (left plot) and cumulative distributions (right plot) of $\hat{s}$ and $\hat{s}^*$ (solid black curve), and $\check{s}$ (dashed grey curve), and $\tilde{s}$ (dashed-dotted grey curve), along with the standard-normal density (solid grey curve). Plots generated with $\varphi_0=1$ and $(\N,\T)=(10,5)$.}
\label{fig:manynormalmeans}
\end{figure}

To simplify the presentation we use the shorthand notation
$$
\hat{e}\coloneqq \sqrt{\N\T}(\hat{\varphi}-\varphi_0),
\qquad
\hat{s}  \coloneqq 2^{-\nicefrac{1}{2}}\, \hat{e}/\hat{\varphi},
$$
for the (scaled) sampling error of the maximum-likelihood estimator and for its studentized version, respectively. The bootstrap quantities $\hat{e}^*$ and $\hat{s}^*$ are defined analogously. We similarly let
$$
\check{e}\coloneqq \sqrt{\N\T}(\check{\varphi}-\varphi_0),
\qquad
\check{s}  \coloneqq 2^{-\nicefrac{1}{2}}\, \check{e}/\hat{\varphi},
\qquad
\tilde{s}  \coloneqq 2^{-\nicefrac{1}{2}}\, \check{e}/\check{\varphi},
$$
for the bias-corrected estimator. The upper-left and upper-right plots in Figure \ref{fig:manynormalmeans} contain, respectively, the density and distribution functions of these quantities for $(n,m)=(10,5)$ and $\varphi_0=1$. The solid black curves refer to $\hat{e}$. The dashed black curves capture the behavior of $\hat{e}^*$ up to first order (i.e., by setting $\epsilon=0$, thereby ignoring the randomness induced by its dependence on the original sample). The solid grey curves, in turn, refer to a mean-zero normal variable with variance $2\varphi_0$ while the dashed grey curves depict, $\check{e}$, the analytically bias-corrected estimator. Here, the distribution of $\hat{e}^*$ does not have quite enough mass in the left tail, compared to the distribution of $\hat{e}$, but mimics the right-tail well. The sampling distribution of $\check{e}$, compared to that of $\hat{e}$, is closer to the normal reference distribution but the sample size is not sufficiently large for the distribution to resemble well its normal approximation. The lower plots in Figure \ref{fig:manynormalmeans} provide corresponding results for the studentized estimators. All these distributions are pivotal and, hence, independent of $\varphi_0$. Here, $\hat{s}$ and $\hat{s}^*$ follow exactly the same distribution; it is given by the solid black curve. The dashed grey curves for $\check{s}$ are the same as those for $\hat{s}$ (and $\hat{s}^*$) up to a translation that brings them closer to the standard-normal reference curves (in solid grey). The distribution of $\check{s}$ has considerable excess mass in its left tail so that confidence intervals constructed by treating it as standard normal  will be too short. By using an unbiased estimator of the asymptotic variance, $\tilde{s}$ reduces this issue somewhat and yields a sampling distribution that is closer to the normal benchmark.

\begin{table}  
\centering
\begin{tabular}{cc|ccccc}
	\hline\hline
 $\N$ & $\T$ &  $\hat{s}$ & $\check{s}$ & $\tilde{s}$ &  $\hat{e}^*$ & $\hat{s}^*$   \\	

 10 & 10 &  0.765  &  0.871  &  0.897  &   0.918 & 0.950 \\

 20 & 10 &  0.682  &  0.868  &  0.897  &   0.918 & 0.950 \\

 40 & 10 &  0.535  &  0.864  &  0.894  &   0.916 & 0.950 \\

100 & 10 &  0.235  &  0.854  &  0.887  &   0.911 & 0.950 \\
\hline\hline
\end{tabular}	
\caption{Many normal means: Coverage of two-sided $95$\% confidence intervals} \label{table:table:manynormalmeans}
\end{table}

To complement this graphical illustration, Table \ref{table:table:manynormalmeans} gives coverage rates of two-sided $95\%$ confidence intervals for $\varphi_0$ across different sample sizes. These rates are invariant to the value of $\varphi_0$. The conclusions from the graphical analysis are borne out in the table. Moreover, the naive normal approximation does poorly when applied to maximum likelihood but bootstrapping the maximum-likelihood estimator yields reliable inference. Here, the percentile $t$-method gives exact coverage but this will not be true in general. Both bootstrap procedures perform better in terms of coverage as those based on bias correction. The table also confirms the relative improvement of $\tilde{s}$ over $\check{s}$. We note, though, that the construction of an unbiased estimator of the sampling variance of $\hat{\varphi}$ is specific to this example. In general, an estimator of the Fisher information will have bias of order $\nicefrac{1}{\T}$. This would be so even if it were possible to use the true common parameter value in the construction of the estimator of $\varSigma$, as the Fisher information also depends on the fixed effects. Replacing them with their estimators causes bias that is of order $\nicefrac{1}{\T}$.

\paragraph{Dynamic logit}
For our next example we consider the Markov process
$$
y_{it} = 
\left\lbrace
\begin{array}{cl}
1 & \text{ if } \eta_{i0} + \varphi_0 y_{it-1} > \varepsilon_{it} \\
0 & \text{ if not}	
\end{array}	
\right. ,
$$
where the $\varepsilon_{it}$ are independent and identically distributed logistic random variables, i.e., $\mathbb{P}(\varepsilon_{it}\leq a) = (1+e^{-a})^{-1}\eqqcolon F(a)$. The initial conditions, $y_{i0}$, are observed and held fixed throughout.

In this example the maximum-likelihood estimator is not available in closed form. Nonetheless, the log-likelihood function is globally concave and numerical optimization via a Newton-Raphson procedure is straightforward (see the next section for details). 
Given $\hat{\varphi}$ and $\hat{\eta}_1,\ldots,\hat{\eta}_n$ we generate bootstrap samples by recursively drawing $y_{it}^*$ from a Bernoulli distribution with success probability $F(\hat{\eta}_i + \hat{\varphi}y_{it-1}^*)$. 


The exact distribution of $\hat{\varphi}$ is not known so we resort to simulations. We draw $y_{i0}$ from its stationary distribution,
$$
\mathbb{P}(y_{i0}=1) = 
\frac{F(\eta_{i0})}{1-F(\eta_{i0}+\varphi_0)+F(\eta_{i0})},
$$
set $\eta_{i0}=0$ for all the strata, and consider $\varphi_0\in \lbrace \nicefrac{1}{2},1 \rbrace$. Table \ref{table:table:logit} provides the coverage rate of (two-sided) 95\% confidence intervals for the autoregressive parameter together with their average length. Results are reported for confidence intervals based on maximum likelihood ($\hat{s}$), on the percentile and percentile $t$-bootstrap ($\hat{e}^*$ and $\hat{s}^*$, respectively), as well as on two procedures that adjust the maximum-likelihood estimator for its bias. The first adjustment ($\check{s}$) is the analytical correction of \cite{HahnKuersteiner2011}. The second adjustment ($\doublecheck{s}$) is due to \cite{FernandezVal2009} and exploits the model structure to implement a refined correction that replaces certain sample averages by expected quantities. Both these approaches require a bandwidth choice. We report results for a bandwidth equal to one, which we found was the choice that performed best here. The bootstrap results, in turn, are based on the use of 999 bootstrap replications. The results in the table are based on 5,000 Monte Carlo replications.

\begin{table} 
\centering
\resizebox{\columnwidth}{!}{
\begin{tabular}{ccc|ccccc|ccccc}
	\hline\hline
\multicolumn{3}{c|}{} & \multicolumn{5}{c|}{COVERAGE} & \multicolumn{5}{c}{LENGTH} \\
$\varphi_0$ & $\N$ & $\T$ & 
 $\hat{s}$ & $\check{s}$ & $\doublecheck{s}$ &  $\hat{e}^*$ & $\hat{s}^*$ & 
 $\hat{s}$ & $\check{s}$ & $\doublecheck{s}$ &  $\hat{e}^*$ & $\hat{s}^*$  \\	

$\nicefrac{1}{2}$ & 100 & 10 &  
0.117 &  0.940 & 0.970 & 0.970 & 0.930 &
0.567 &  0.572 & 0.574 & 0.629 & 0.542 \\

$\nicefrac{1}{2}$ & 100 & 20 &
0.381  &  0.958  &  0.965  &  0.956  &  0.951 &
0.378  &  0.381  &  0.381  &  0.395  &  0.372 \\

$\nicefrac{1}{2}$ & 250 & 10 &  	
0.001 &  0.887 & 0.953 & 0.963 & 0.907 & 
0.358 &  0.362 & 0.363 & 0.397 & 0.344 \\

$\nicefrac{1}{2}$ & 250 & 20 & 
0.046  &  0.932  &  0.949  &  0.952  &  0.943 &
0.239  &  0.241  &  0.241  &  0.250  &  0.236 \\

1 & 100 & 10 &  
0.095 &  0.878 & 0.933 & 0.957 & 0.907 & 
0.605 &  0.620 & 0.623 & 0.656 & 0.577 \\

1 & 100 & 20 & 
0.329  &  0.921  &  0.944  &  0.953  &  0.944 &
0.404  &  0.410  &  0.410  &  0.418  &  0.398 \\

1 & 250 & 10 & 
0.001 &   0.699 &   0.891 &   0.955 &   0.893 &
0.383 &   0.392 &   0.394 &   0.413 &   0.365 \\

1 & 250 & 20 & 
0.027  &  0.866  &  0.910  &  0.965  &  0.943 &
0.255  &  0.259  &  0.259  &  0.264  &  0.252 \\

\hline\hline

\end{tabular}	
}	
\caption{Dynamic logit: Properties of two-sided $95\%$ confidence intervals} \label{table:table:logit}
\end{table}

The naive normal approximation to the sampling distribution of the maximum-likelihood estimator again yields unreliable inference in this problem. Bias correction yields a large improvement in coverage rates and comes with only minor increases in the length of the confidence intervals (which is informative about efficiency). Confidence intervals based on the correction underlying $\doublecheck{s}$ tend to give better coverage than those based on $\check{s}$, with the difference sometimes being considerable (up to $20$ percentage points). This highlights the sensitivity of bias-corrected inference to how the bias is being estimated; this is an issue not accounted for by first-order theory. The bootstrap, rather than estimating the bias, mimics it. Both $\hat{e}^*$ and $\hat{s}^*$ are competitive with bias correction, doing at least as well as $\doublecheck{s}$ in terms of coverage. 
The percentile $t$-method yields shorter confidence intervals but, for $\T=10$, this comes at the cost of some undercoverage. This problem is essentially resolved for $\T=20$. 

\paragraph{Many normal means (cont'd)}
In our third and final example we reconsider the setup of \cite{NeymanScott1948} but change the parameter of interest to
$$
\Delta = \lim_{\N\rightarrow\infty} \frac{1}{\N} \sum_{i=1}^\N \eta_{i0}^2,
$$
the second moment of the fixed effects. 
The plug-in estimator is
$\nicefrac{1}{\N} \sum_{i=1}^\N \overline{z}_i^2$. 
Using the fact that $\overline{z}_i\sim N(\eta_{i0},\nicefrac{\varphi_0}{\T})$ by normality of the data it is easy to verify that the plug-in bias due to the estimation of the fixed effects is $\nicefrac{\varphi_0}{\T}$, while the estimator's sampling variance is 
$$
\frac{2\varphi_0}{\N\T} \left( 2 \frac{\sum_{i=1}^\N \eta_{i0}^2}{\N} + \frac{\varphi_0}{\T} \right).
$$
The second component in the expression of the variance is of smaller order and not picked up by our general expression for $\sigma^2$ given previously.

The exact distribution of the estimator is a complicated mixture and so we again resort to simulations to evaluate the performance of the bootstrap.
In our simulations we set $\eta_{i0} = i/\N$ so that, in large samples, the distribution of the fixed effects is uniform on $[0,1]$; hence, $\Delta = \nicefrac{1}{3}$. Data were generated with $\varphi_0 = 1$. We report results for several choices of $(\N,\T)$ in Table \ref{table:normalmeans2}. The bootstrap confidence intervals are  again found to yield a large improvement in coverage relative to the ones based on the naive plug-in approach. Again the simple percentile method does slightly better than the percentile-$t$ method. The average length of the former's confidence intervals co-incide (up to the fourth decimal digit) with those of maximum likelihood.

\begin{table}  
\centering
\begin{tabular}{cc|ccc|ccc}
	\hline\hline
\multicolumn{2}{c|}{} & \multicolumn{3}{c|}{COVERAGE} & \multicolumn{3}{c}{LENGTH} \\
$\N$ & $\T$ & 
$\hat{s}$ &  $\hat{e}^*$ & $\hat{s}^*$ & 
$\hat{s}$ &  $\hat{e}^*$ & $\hat{s}^*$ 
 \\

50   & 10 &  0.545  &  0.945  &  0.917  &  0.232  &  0.232  &  0.210 \\
50   & 20 &  0.704  &  0.958  &  0.934  &  0.156  &  0.156  &  0.145 \\
50   & 50 &  0.787  &  0.946  &  0.926  &  0.095  &  0.095  &  0.091 \\
100  & 10 &  0.258  &  0.969  &  0.924  &  0.164  &  0.163  &  0.147 \\
100  & 20 &  0.490  &  0.956  &  0.933  &  0.110  &  0.110  &  0.102 \\
100  & 50 &  0.718  &  0.935  &  0.921  &  0.067  &  0.067  &  0.064 \\

\hline\hline

\end{tabular}	
\caption{Many normal means: Properties of two-sided $95\%$ confidence intervals for $\lim_{\N\rightarrow\infty} \nicefrac{1}{\N} \sum_{i=1}^\N \eta_{i0}^2$} \label{table:normalmeans2}
\end{table}

\section{A note on implementation}
In most applications the bootstrap distribution is unknown and needs to be simulated. This, in turn, requires computation of the maximum-likelihood estimator many times. In spite of the presence of a large number of fixed effects, a standard Newton-Raphson procedure is feasible here by exploiting the sparsity of the Hessian matrix. Furthermore, as many popular fixed-effect specifications involve log-likelihood functions that are globally concave, such an algorithm is numerically stable and requires only few iterations to locate the global maximizer.

Collect all parameters in $\theta\coloneqq(\varphi,\eta_1,\ldots,\eta_\N)$. A Newton step starting at $\theta$ is of the form
$$
\theta - \ell_{\theta\theta}^{-1} \, \ell_{\theta}^{\vphantom{-1}},
$$
where $\ell_\theta$ and $\ell_{\theta\theta}$ are the score vector and Hessian matrix. The Hessian matrix is large and so direct inversion can be both slow and numerically inaccurate. Fortunately, the Hessian has a particular block structure. Moreover, 
$$
\ell_\theta = 
\left(\begin{array}{c} \ell_\varphi \\ \ell_{\eta_1} \\ \ell_{\eta_2} \\ \vdots \\ \ell_{\eta_\N} \end{array}\right)
\qquad
\ell_{\theta\theta} = 
\left(
\begin{array}{ccccc}
\ell_{\varphi\varphi} & \ell_{\varphi\eta_1}  & \ell_{\varphi\eta_2} & \cdots & \ell_{\varphi\eta_\N} \\ 	
\ell_{\eta_1\varphi}  & \ell_{\eta_1\eta_1} & 0  & \cdots & 0 \\
\ell_{\eta_2\varphi}  & 0             & \ell_{\eta_2\eta_2}  & \ddots & 0 \\
\vdots & \vdots & \ddots & \ddots & \vdots \\
\ell_{\eta_\N\varphi}  & 0         & 0 & \cdots & \ell_{\eta_\N\eta_\N} \\
\end{array}	
\right),
$$
where the individual components are 
\begin{equation*}
\begin{aligned}
\ell_{\varphi} & \coloneqq
\sum_{i=1}^\N \sum_{t=1}^\T
\frac{\partial \ell(\varphi,\eta_i \vert z_{it})}{\partial \varphi} \, \, ,	\qquad
\\
\ell_{\varphi\varphi} & \coloneqq
\sum_{i=1}^\N \sum_{t=1}^\T 
\frac{\partial^2 \ell(\varphi,\eta_i \vert z_{it})}{\partial \varphi\partial\varphi^\prime},
\end{aligned}
\begin{aligned}
\ell_{\eta_i} 
& \coloneqq
\sum_{t=1}^\T
\frac{\partial \ell(\varphi,\eta_i \vert z_{it})}{\partial \eta_i} \, \, , \qquad
\\
\ell_{\eta_i\eta_i} & \coloneqq
\sum_{t=1}^\T 
\frac{\partial^2 \ell(\varphi,\eta_i \vert z_{it})}{\partial \eta_i \partial\eta_i^\prime},
\end{aligned}
\end{equation*}
and
$$
\ell_{\varphi\eta_i}\coloneqq
\sum_{t=1}^\T 
\frac{\partial^2 \ell(\varphi,\eta_i \vert z_{it})}{\partial \varphi \partial\eta_i^\prime} = \ell_{\eta_i\varphi}^\prime.
$$
By making use of partitioned-invere formulae we arrive at an expression for $\ell_{\theta\theta}^{-1}$ that can be computed by using only the inverses of the substantially smaller matrices $\ell_{\varphi\varphi}$ and $\ell_{\eta_i\eta_i}$. With 

$$
\ell_{\theta\theta}^{-1} = 
\left(
\begin{array}{ccccc}
(\ell_{\theta\theta}^{-1})_{\varphi\varphi} & (\ell_{\theta\theta}^{-1})_{\varphi\eta_1}  & (\ell_{\theta\theta}^{-1})_{\varphi\eta_2} & \cdots & (\ell_{\theta\theta}^{-1})_{\varphi\eta_\N} \\ 	
(\ell_{\theta\theta}^{-1})_{\eta_1\varphi}  & (\ell_{\theta\theta}^{-1})_{\eta_1\eta_1} & (\ell_{\theta\theta}^{-1})_{\eta_1\eta_2}  & \cdots & (\ell_{\theta\theta}^{-1})_{\eta_1\eta_\N} \\
(\ell_{\theta\theta}^{-1})_{\eta_2\varphi}  & (\ell_{\theta\theta}^{-1})_{\eta_2\eta_1}             & (\ell_{\theta\theta}^{-1})_{\eta_2\eta_2}  & \ddots & (\ell_{\theta\theta}^{-1})_{\eta_2\eta_\N} \\
\vdots & \vdots & \ddots & \ddots & \vdots \\
(\ell_{\theta\theta}^{-1})_{\eta_\N\varphi}  & (\ell_{\theta\theta}^{-1})_{\eta_\N\eta_1}         & (\ell_{\theta\theta}^{-1})_{\eta_\N\eta_2}  & \cdots & (\ell_{\theta\theta}^{-1})_{\eta_\N\eta_\N} \\
\end{array}	
\right),
$$
we have
$$
(\ell_{\theta\theta}^{-1})_{\varphi\varphi}
\coloneqq
\left( 
\ell_{\varphi\varphi}^{\vphantom{-1}} - \sum_{i=1}^n \ell_{\varphi\eta_i}^{\vphantom{-1}} \, \ell_{\eta_i\eta_i}^{-1} \, \ell_{\eta_i\varphi}^{\vphantom{-1}}
\right)^{-1}, 
\qquad
(\ell_{\theta\theta}^{-1})_{\varphi\eta_i}
\coloneqq
-
(\ell_{\theta\theta}^{-1})_{\varphi\varphi} 
\ell_{\varphi\eta_i}^{\vphantom{-1}}
\ell_{\eta_i\eta_i}^{-1}
=
(\ell_{\theta\theta}^{-1})_{\eta_i\varphi}^\prime,
$$
and, treating the cases where $i=j$ and $i\neq j$ separately for clarity, 
$$
(\ell_{\theta\theta}^{-1})_{\eta_i\eta_i}
 \coloneqq
\ell_{\eta_i\eta_i}^{-1} 
+ 
\ell_{\eta_i\eta_i}^{-1} \,
\ell_{\eta_i\varphi}^{\vphantom{-1}} \,
(\ell^{-1}_{\theta\theta})_{\varphi\varphi} \,
\ell_{\varphi\eta_i}^{\vphantom{-1}} \,
\ell_{\eta_i\eta_i}^{-1}
\qquad
(\ell_{\theta\theta}^{-1})_{\eta_i\eta_j}
 \coloneqq
\ell_{\eta_i\eta_i}^{-1} \,
\ell_{\eta_i\varphi}^{\vphantom{-1}} \,
(\ell^{-1}_{\theta\theta})_{\varphi\varphi} \,
\ell_{\varphi\eta_j}^{\vphantom{-1}} \,
\ell_{\eta_j\eta_j}^{-1}.
$$
The Newton step for $\varphi$ then simply is
$$
\varphi - 
(\ell_{\theta\theta}^{-1})_{\varphi\varphi} \, \ell_\varphi
-
\sum_{i=1}^\N
(\ell_{\theta\theta}^{-1})_{\varphi\eta_i} \, \ell_{\eta_i}
=
\varphi - 
(\ell_{\theta\theta}^{-1})_{\varphi\varphi}^{\vphantom{-1}}	\,\left( 
\ell_\varphi^{\vphantom{-1}}
-
\sum_{i=1}^\N \ell_{\varphi\eta_i}^{\vphantom{-1}} \, \ell_{\eta_i\eta_i}^{-1} \, \ell_{\eta_i}^{\vphantom{-1}}
\right).
$$
The corresponding step for each fixed effect $\eta_i$ is
$$
\eta_i - 
(\ell_{\theta\theta}^{-1})_{\eta_i\varphi} \, \ell_{\varphi}
-
\sum_{j=1}^\N
(\ell_{\theta\theta}^{-1})_{\eta_i\eta_j} \, \ell_{\eta_j}
=
\eta_i - 
\ell_{\eta_i\eta_i}^{-1}
\left( 
\ell_{\eta_i}^{\vphantom{-1}} 
- 
 \ell_{\eta_i\varphi}^{\vphantom{-1}} 
(\ell_{\theta\theta}^{-1})_{\varphi\varphi}^{\vphantom{-1}}	\,
\left(  \ell_\varphi^{\vphantom{-1}}
-
\sum_{j=1}^\N \ell_{\varphi\eta_{j}}^{\vphantom{-1}} \, \ell_{\eta_{j}\eta_{j}}^{-1} \, \ell_{\eta_{j}}^{\vphantom{-1}}
\right)
\right)
.
$$
A Newton-Raphson algorithm that uses these updating formulae is feasible even in large data sets. The size of the matrices to be inverted is independent of the sample size. The computational complexity is, therefore, comparable to that of the setting without fixed effects.

\section{Asymptotic theory}

Our results hold under a set of assumptions that are standard in the literature. The following formulation is mostly borrowed from \cite{KimSun2016}. It differs from \cite{HahnKuersteiner2011} in two respects that are worth noting. The first difference is that the individual time series need not be stationary. This is useful because the requirement that the initial condition is a draw from the steady-state distribution, for example, is often hard to justify. 
The second difference is that certain requirements are assumed to hold uniformly over a neighborhood of the true parameter value. This is useful for the derivation of our results because, like \cite{KimSun2016}, we adopt a technique introduced in \cite{Andrews2005} to obtain these.
This technique is to first demonstrate a convergence result for the maximum-likelihood estimator uniformly over a set around the true parameter value. Then, as consistency implies that the maximum-likelihood estimator lies in this set with probability approaching one, this allows us the establish the corresponding property for the bootstrap estimator.

In the assumptions (and in the proofs) it is important to make clear under which data generating process certain expectations and probabilities are being computed. We will write $\mathbb{E}_\theta$ and $\mathbb{P}_\theta$ for expectations and probabilities involving data that were generated using parameters $\theta=(\varphi,\eta_1,\ldots,\eta_\N)$. Note that some objects, such as $\mathbb{E}_\theta(z_{it})$, only depend on a subset of the elements of $\theta$. For simplicity, however, we do not make this explicit in the notation.

Denote by $V_\varphi$ and $V_\eta$ the parameter space for $\varphi$ and $\eta_i$, respectively. Then the parameter space for $\theta$ is the Cartesian product $\varTheta\coloneqq V_\varphi \times V_\eta \times \cdots \times V_\eta$. We let $\varTheta_0$ be a subset of $\varTheta$. 

\begin{assumption} \label{ass:ass1}
\mbox{} \newline
(i) The function $f$ is continuous in $\varphi\in V_\varphi$ and $\eta_i\in V_\eta$. 

\medskip\noindent
(ii) The true parameter value lies in the interior of $\varTheta_0$, a subset of the compact set $\varTheta$.
\end{assumption}

For our next assumption, consider the mixing coefficients 
$$
a_i(\theta,h) \coloneqq
\sup_{1\leq t \leq \T} 
\sup_{A\in\mathcal{A}_{it}(\theta)}
\sup_{B\in \mathcal{B}_{it+h}(\theta)}
\lvert
\mathbb{P}_{\theta}(A\cap B)
-
\mathbb{P}_{\theta}(A) \, \mathbb{P}_{\theta}(B)  
\rvert,
$$
where $\mathcal{A}_{it}(\theta)$ and $\mathcal{B}_{it}(\theta)$ are the sigma algebras generated by the sequences $z_{it},z_{it-1},\ldots$ and $z_{it},z_{it+1},\ldots$ when these sequences were generated from our model with the parameter equal to $\theta$. 

We will also make use of an open set that covers $\varTheta_0$. This set is of the form
$$
\varTheta_1 \coloneqq \lbrace  \theta\in \varTheta : d(\theta,\varTheta_0) < \delta \rbrace
$$
for some $\delta>0$, where $d(\theta,\varTheta_0)\coloneqq\inf \lbrace \lVert \theta-\vartheta \rVert_2 : \vartheta\in\varTheta_0 \rbrace$, i.e., the distance between the point $\theta$ and the set $\varTheta_0$.

\begin{assumption} \label{ass:ass2}
$
\sup_{1\leq i\leq \N} \sup_{\theta\in \varTheta_1} a_i(\theta,h) 
=
O(r^h)
$
for some constant $0<r<1$.
\end{assumption}

The next assumption collects smoothness conditions and moment requirements.

\begin{assumption} \label{ass:ass3}
\mbox{} \newline
(i) The function $\ell(\varphi,\eta_i \vert z_{it})$ is four times continuously-differentiable in $\varphi$ and $\eta_i$.

\medskip\noindent
(ii) All cross-derivatives of $\ell(\varphi,\eta_i \vert z_{it})$ up to fourth order are bounded by a function $b(z_{it})$ for which 
$$
\sup_{1\leq i \leq \N} \sup_{1\leq t \leq \T} \sup_{\theta\in\Theta_1} \mathbb{E}_\theta(\lvert b(z_{it}) \rvert^{p}) < \infty
$$
for some $p> \dim \varphi+\dim\eta_i+11$.

\medskip\noindent
(iii) As $\T\rightarrow\infty$, 
$\nicefrac{1}{\T} \sum_{t=1}^\T \mathbb{E}_\theta (b(z_{it}))$ converges to 
$\lim_{\T\rightarrow\infty}\nicefrac{1}{\T} \sum_{t=1}^\T \mathbb{E}_\theta (b(z_{it}))$ uniformly in $i$ and $\theta\in\varTheta_1$.
\end{assumption}

Let
$$
G_i(\varphi,\eta_i\vert \vartheta) \coloneqq
\lim_{\T\rightarrow\infty} \frac{1}{\T} \sum_{t=1}^\T
\mathbb{E}_\vartheta(\ell(\varphi,\eta_i \vert z_{it})).
$$
The next assumption ensures that our parameters are identified from time series variation.

\begin{assumption} \label{ass:ass4}
For each $\varepsilon>0$ 	
$$
\inf_{1\leq i \leq \N}
\inf_{\theta\in\varTheta_1}
\left( 
G_i(\varphi,\eta_i \vert \theta) \ 
-
\sup_{\lbrace (\bar{\varphi},\bar{\eta}_i):\lVert (\bar{\varphi}, \bar{\eta}_i) - (\varphi,\eta_i)  \rVert_2 > \varepsilon \rbrace}
G_i( \bar{\varphi},\bar{\eta}_i \vert \theta)
\right)	
>0 .
$$
\end{assumption}

Assumption \ref{ass:ass5} states that we are working under rectangular-array asymptotics.

\begin{assumption} \label{ass:ass5}
As $\N,\T\rightarrow \infty$, $\nicefrac{\N}{\T}\rightarrow \gamma^2$ for some $0<\gamma<\infty$.
\end{assumption}

The final assumption ensures a well-defined asymptotic variance matrix for $\hat{\varphi}$. 

\begin{assumption} \label{ass:ass6}
There exist non-zero finite constants $\epsilon_1,\epsilon_2$ and $\varepsilon_1,\varepsilon_2$  such that, for $\N$ and $\T$ large enough, 

\noindent
(i) \vspace{-1.20cm}
\begin{equation*}
\begin{split}	
&
\inf_{1\leq i\leq n} \inf_{\theta\in\varTheta_1} 
\mathrm{mineig}
\left(
\frac{1}{\T} \sum_{t=1}^\T \mathbb{E}_\theta\left(\frac{\partial^2 \ell (\varphi,\eta_i\vert z_{it})}{\partial\eta_i\partial\eta_i^\prime}\right)
\right) 
\geq \epsilon_1,
\\
&
\sup_{1\leq i\leq n} \sup_{\theta\in\varTheta_1} \left\lVert \frac{1}{\T} \sum_{t=1}^\T \mathbb{E}_\theta\left(\frac{\partial^2 \ell (\varphi,\eta_i\vert z_{it})}{\partial\eta_i\partial\varphi^\prime}\right) \right\rVert_2 \leq \epsilon_2, \text{ and }
\end{split}
\end{equation*}

\medskip\noindent	
(ii) $\varepsilon_1 < \inf_{\theta\in\varTheta_1}\lambda_{\min}(\theta) \leq \sup_{\theta\in\varTheta_1}\lambda_{\max}(\theta) < \varepsilon_2$, where $\lambda_{\min}(\theta)$ and $\lambda_{\max}(\theta)$ are the smallest and largest eigenvalue of the sample average of
$$
-
\mathbb{E}_\theta
\left( 
\frac{\partial^2 \ell (\varphi,\eta_i\vert z_{it})}{\partial\varphi\partial\varphi^\prime}
-
\mathbb{E}_\theta
\left(\sum_{t=1}^\T\frac{\partial^2 \ell (\varphi,\eta_i\vert z_{it})}{\partial\varphi\partial\eta_i^\prime}\right) \,
\mathbb{E}_\theta
\left(\sum_{t=1}^\T\frac{\partial^2 \ell (\varphi,\eta_i\vert z_{it})}{\partial\eta_i\partial\eta_i^\prime}\right)
^{-1}
\frac{\partial^2 \ell (\varphi,\eta_i\vert z_{it})}{\partial\eta_i\partial\varphi^\prime}
\right),
$$
the Fisher information on $\varphi$ contained in observation $z_{it}$.
\end{assumption}

Our main result is stated in the following theorem.

\begin{theorem} \label{thm:thm1}
Let Assumptions \ref{ass:ass1}--\ref{ass:ass6} hold. Then
$$
\mathbb{P}
\left(
\sup_a
\left\lvert
\mathbb{P}^*(\sqrt{\N\T}(\hat{\varphi}^*-\hat{\varphi})\leq a)
-
\mathbb{P}(\sqrt{\N\T}(\hat{\varphi}-\varphi)\leq a)
\right\rvert
> \varepsilon
\right)
=
o(1)
$$	
for any $\varepsilon>0$.
\end{theorem}	

\noindent
Theorem \ref{thm:thm1} justifies the use of the percentile bootstrap for inference.

Next, let 
$$
\hat{\varSigma}\coloneqq
-
\left( 
\frac{1}{\N\T} \sum_{i=1}^\N \sum_{t=1}^\T
\left(
\frac{\partial^2 \ell (\hat{\varphi},\hat{\eta}_i\vert z_{it})}{\partial\varphi\partial\varphi^\prime}
-
\hat{\rho}_i 
\frac{\partial^2 \ell (\hat{\varphi},\hat{\eta}_i\vert z_{it})}{\partial\eta_i\partial\varphi^\prime}
\right)
\right)
^{-1}
$$
be the plug-in estimator of $\varSigma$ based on the maximum-likelihood estimator, 
where we let
$$
\hat{\rho}_i
\coloneqq
\left(
\frac{1}{\T}
\sum_{t=1}^\T
\frac{\partial^2 \ell (\hat{\varphi},\hat{\eta}_i\vert z_{it})}{\partial\varphi\partial\eta_i^\prime}\right) \,
\left(
\frac{1}{\T}
\sum_{t=1}^\T
\frac{\partial^2 \ell (\hat{\varphi},\hat{\eta}_i\vert z_{it})}{\partial\eta_i\partial\eta_i^\prime}\right)
^{-1}.
$$
A consistency result for this estimator, as well as for its bootstrap counterpart, is given next.

\begin{theorem} \label{thm:thm2}
Let Assumptions \ref{ass:ass1}--\ref{ass:ass6} hold. Then
$
\hat{\varSigma} \overset{P}{\rightarrow} \varSigma
$
and
$
\hat{\varSigma}^* \overset{P^*}{\rightarrow} \varSigma.
$
\end{theorem}

\noindent
Both results, when taken together, justify an application of the bootstrap to standardized quantities such as the Wald statistic, for example.

\section*{Conclusion}
The purpose of this paper has been to show that in panel data models with fixed effects, inference based on the bootstrap remains valid under rectangular-array asymptotics. Our results cover quite general nonlinear models and allow for dynamics in the outcome of interest. 

The main advantage of the bootstrap is that it avoids the need to correct for the asymptotic bias in the limit distribution of the maximum-likelihood estimator. It is unlikely that, in our context, the bootstrap yields asymptotic refinements in general as the presence of bias renders the limit distribution non-pivotal, even after studentization. It could be of interest to investigate whether refinements can be obtained by combining the bootstrap with bias correction. On the other hand, the parametric bootstrap we consider is restricted to the correctly-specified likelihood setting. While this is arguably the default for nonlinear panel problems, some of the approaches to bias correction can be generalized to other settings, such as partial likelihoods. In related work, \cite{GoncalvesKaffo2015} have shown that a version of the wild bootstrap replicates the bias in the setup of \cite{HahnKuersteiner2002}. However, their approach is residual-based and is tailored quite specifically to the linear model.

While our attention has been devoted to one-way models, we see no reason why our main message would not carry over to models with two-way fixed effects.
The available results on the behavior of the maximum-likelihood estimator of such models are more restrictive, however, in that they impose additive or multiplicative restrictions on the way the fixed effects enter the likelihood; see \cite{FernandezValWeidner2016} for bias expressions (and corrections) in such a setting.

\appendix
\renewcommand{\theequation}{A.\arabic{equation}} 
\setcounter{equation}{0}
\renewcommand{\thelemma}{A.\arabic{lemma}} 
\setcounter{lemma}{0}
\section*{Appendix} 

\paragraph{Proof of Theorem \ref{thm:thm1}.}
Note that
$$
\mathbb{P}
\left(
\sup_a
\left\lvert
\mathbb{P}^*(\sqrt{\N\T}(\hat{\varphi}^*-\hat{\varphi})\leq a)
-
\mathbb{P}(\sqrt{\N\T}(\hat{\varphi}-\varphi_0)\leq a)
\right\rvert
> \varepsilon
\right)
$$
is bounded from above by 
$$
\sup_{\theta\in\varTheta_0}
\mathbb{P}_\theta
\left(
\sup_a
\left\lvert
\mathbb{P}_{\hat{\theta}}(\sqrt{\N\T}(\hat{\varphi}^*-\hat{\varphi})\leq a)
-
\mathbb{P}_\theta(\sqrt{\N\T}(\hat{\varphi}-\varphi)\leq a)
\right\rvert
> \varepsilon
\right)
$$
which, in turn, is below
\begin{equation} \label{eq:bound1}
\begin{split}	
&
\sup_{\theta\in\varTheta_0}
\mathbb{P}_\theta
\left(
\sup_a
\left\lvert
\mathbb{P}_\theta(\sqrt{\N\T}(\hat{\varphi}^{\hphantom{*}}-\varphi)\leq a)
-
\mathbb{P}_\theta ( v_\theta \leq a )
\right\rvert
> \frac{\varepsilon}{2}
\right) 
\\
+ &
\sup_{\theta\in\varTheta_0}
\mathbb{P}_\theta
\left(
\sup_a
\left\lvert
\mathbb{P}_{\hat{\theta}}(\sqrt{\N\T}(\hat{\varphi}^*-\hat{\varphi})\leq a)
-
\mathbb{P}_\theta ( v_\theta \leq a )
\right\rvert
> \frac{\varepsilon}{2}
\right)
.
\end{split}
\end{equation}
Here and later, we let 
$$
v_\theta\sim N(\gamma \beta_\theta,\varSigma_\theta)
$$ 
for $\beta_\theta$ and $\varSigma_\theta$ the asymptotic bias and asymptotic variance of the maximum-likelihood estimator for data generated with parameter $\theta$. Therefore, it suffices to show that each of the terms in \eqref{eq:bound1} is $o(1)$.

First, starting from Theorem 2 in \cite{KimSun2016} we obtain (see the supplement) 
$$
\sup_{\theta\in\varTheta_1}
\left\lvert
\mathbb{P}_\theta(\sqrt{\N\T}(\hat{\varphi}^{\hphantom{*}}-\varphi)\leq a)
-
\mathbb{P}_\theta ( v_\theta \leq a )
\right\rvert
=
o(1)
$$
for any $a$. Further, because the normal distribution is a continuous function, we have that
\begin{equation} \label{eq:andrews1}
\sup_{\theta\in\varTheta_1}
\left(  \sup_a
\left\lvert
\mathbb{P}_\theta(\sqrt{\N\T}(\hat{\varphi}^{\hphantom{*}}-\varphi)\leq a)
-
\mathbb{P}_\theta ( v_\theta \leq a )
\right\rvert
\right) 
=
o(1)
\end{equation}
by Polya's theorem. This allows us to envoke Lemma A.1~of \cite{Andrews2005} to establish that
$$
\sup_{\theta\in\varTheta_0} \mathbb{P}_\theta
\left( 
\sup_a
\left\lvert
\mathbb{P}_\theta(\sqrt{\N\T}(\hat{\varphi}^{\hphantom{*}}-\varphi)\leq a)
-
\mathbb{P}_\theta ( v_\theta \leq a )
\right\rvert
> \frac{\varepsilon}{2}
\right) = o(1).
$$
This handles the first term in \eqref{eq:bound1}.

Moving on to the second term in \eqref{eq:bound1}, note that
\begin{equation*}
\begin{split}
&
\sup_{\theta\in\varTheta_0}
\mathbb{P}_\theta
\left(
\sup_a
\left\lvert
\mathbb{P}_{\hat{\theta}}(\sqrt{\N\T}(\hat{\varphi}^*-\hat{\varphi})\leq a)
-
\mathbb{P}_\theta ( v_\theta \leq a )
\right\rvert
> \frac{\varepsilon}{2}
\right)	
\\
\leq &
\sup_{\theta\in\varTheta_0}
\mathbb{P}_\theta
\left(
\sup_a
\left\lvert
\mathbb{P}_{\hat{\theta}}(\sqrt{\N\T}(\hat{\varphi}^*-\hat{\varphi})\leq a)
-
\mathbb{P}_{\hat{\theta}}(v_{\hat{\theta}} \leq a)
\right\rvert
> \frac{\varepsilon}{4}
\right)	
\\
+ & 
\sup_{\theta\in\varTheta_0}
\mathbb{P}_\theta
\left(
\sup_a
\left\lvert
\mathbb{P}_{\hat{\theta}}(v_{\hat{\theta}} \leq a)
-
\mathbb{P}_\theta ( v_\theta \leq a )
\right\rvert
> \frac{\varepsilon}{4}
\right)	.
\end{split}
\end{equation*}		
Here, using \eqref{eq:andrews1}, coupled with the consistency result
\begin{equation} \label{eq:andrews2}
\sup_{\theta\in\varTheta_1} \mathbb{P}(\lVert \hat{\theta}-
\theta \rVert_2 > \epsilon) = o(1)
\end{equation}
(which follows from Theorem 1 of \citealt{KimSun2016}), by another application of Lemma A.1~of \cite{Andrews2005}, 
$$
\sup_{\theta\in\varTheta_0} \mathbb{P}_\theta
\left( 
\sup_a
\left\lvert
\mathbb{P}_\theta(\sqrt{\N\T}(\hat{\varphi}^{*}-\hat{\varphi})\leq a)
-
\mathbb{P}_{\hat{\theta}} ( v_{\hat{\theta}} \leq a )
\right\rvert
> \frac{\varepsilon}{4}
\right) = o(1)
$$
while, again using \eqref{eq:andrews2}, 
$$
\sup_{\theta\in\varTheta_0}
\mathbb{P}_\theta
\left(
\sup_a
\left\lvert
\mathbb{P}_{\hat{\theta}}(v_{\hat{\theta}} \leq a)
-
\mathbb{P}_\theta ( v_\theta \leq a )
\right\rvert
> \frac{\varepsilon}{4}
\right)
= o(1)
$$
follows from the continuous mapping theorem. This takes care of the second term in \eqref{eq:bound1} and completes the proof of the theorem. \qed

\paragraph{Proof of Theorem \ref{thm:thm2}.}
We introduce the notational shorthand
$$
V_{it}
\coloneqq
\left(\begin{array}{cc}
V_{it}^{11} & V_{it}^{12} \\
V_{it}^{21} & V_{it}^{22} 
\end{array}\right)
=
\left(\begin{array}{cc}
\frac{\partial^2\ell(\varphi,\eta_i \vert z_{it})}{\partial \varphi \partial \varphi^\prime}
& 
\frac{\partial^2\ell(\varphi,\eta_i \vert z_{it})}{\partial \varphi \partial \eta_i^\prime}
\\
\frac{\partial^2\ell(\varphi,\eta_i \vert z_{it})}{\partial \eta_i \partial \varphi^\prime}
& 
\frac{\partial^2\ell(\varphi,\eta_i \vert z_{it})}{\partial \eta_i \partial \eta_i^\prime}
\end{array}\right),
$$
where the derivatives are evaluated at the parameter values that were used to generate the data. 
In the same manner, we write the plug-in estimator constructed using  $\hat{\varphi},\hat{\eta}_i$ as $\hat{V}_{it}$.
The Fisher information on $\varphi$ in observation $z_{it}$ when the true parameter value is $\theta$ then is
$$
\varOmega_{it,\theta} \coloneqq
-
\left( 
\mathbb{E}_\theta (V_{it}^{11})
-
\mathbb{E}_\theta (V_{it}^{12})
(\mathbb{E}_\theta (V_{it}^{22}))^{-1}
\mathbb{E}_\theta (V_{it}^{21})
\right),
$$
and its plug-in estimator is 
$$
\hat{\varOmega}_{it,\theta} \coloneqq
-
\left( 
\frac{1}{\T} \sum_{t=1}^\T \hat{V}_{it}^{11}
-
\left(\frac{1}{\T} \sum_{t=1}^\T \hat{V}_{it}^{12} \right)
\left(\frac{1}{\T} \sum_{t=1}^\T \hat{V}_{it}^{22}\right)^{-1}
\left(\frac{1}{\T} \sum_{t=1}^\T \hat{V}_{it}^{21}\right)
\right).
$$
To show Theorem \ref{thm:thm2} we establish that, for all $\varepsilon > 0$, 
\begin{equation} \label{eq:omega}
\sup_{\theta\in\varTheta_1}
\mathbb{P}_{\theta}
\left( 
\max_{1\leq i\leq \N}
\left\lVert
\frac{1}{\T} \sum_{t=1}^\T 
(\hat{\varOmega}_{it,\theta}
-
{\varOmega}_{it,\theta})
\right\rVert_2
> \varepsilon
\right)
=
o(1).
\end{equation}
We can then use Lemma A.1~of \cite{Andrews2005} to verify the consistency of both  $\hat{\varSigma}$ and $\hat{\varSigma}^*$ as stated in the theorem.

To show \eqref{eq:omega} it suffices to prove that, for all $
\varepsilon>0$, 
\begin{equation*} 
\begin{split}
& \sup_{\theta\in\varTheta_1}
\mathbb{P}_{\theta}
\left( 
\max_{1\leq i\leq \N}
\left\lVert
\frac{1}{\T} \sum_{t=1}^\T 
(\hat{V}_{it}^{11}
-
\mathbb{E}_\theta(
{V}_{it}^{11}))
\right\rVert_2
> \varepsilon
\right)
=
o(1),
\\
& \sup_{\theta\in\varTheta_1}
\mathbb{P}_{\theta}
\left( 
\max_{1\leq i\leq \N}
\left\lVert
\frac{1}{\T} \sum_{t=1}^\T 
(\hat{V}_{it}^{12}
-
\mathbb{E}_\theta (
{V}_{it}^{12}))
\right\rVert_2
> \varepsilon
\right)
=
o(1),
\\
& \sup_{\theta\in\varTheta_1}
\mathbb{P}_{\theta}
\left( 
\max_{1\leq i\leq \N}
\left\lVert
\frac{1}{\T} \sum_{t=1}^\T 
(\hat{V}_{it}^{22}
-
\mathbb{E}_\theta (
{V}_{it}^{22}))
\right\rVert_2
> \varepsilon
\right)
=
o(1).
\end{split}
\end{equation*}
The proof for each of these terms is similar and so we only provide details for the first of them. 

To begin we note that
$$
\sup_{\theta\in\varTheta_1}
\mathbb{P}_{\theta}
\left( 
\max_{1\leq i\leq \N}
\left\lVert
\frac{1}{\T} \sum_{t=1}^\T 
(\hat{V}_{it}^{11}
-
\mathbb{E}_\theta(
{V}_{it}^{11}))
\right\rVert_2
> \varepsilon
\right)	
$$
is bounded from above by 
\begin{equation*}
\begin{split}
\sup_{\theta\in\varTheta_1} \hspace{-.1cm}
\mathbb{P}_{\theta} \hspace{-.1cm}
\left( 
\max_{1\leq i\leq \N}
\left\lVert
\frac{1}{\T} \sum_{t=1}^\T 
(\hat{V}_{it}^{11}
-
{V}_{it}^{11})
\right\rVert_2
> \frac{\varepsilon}{2}
\right)	
\hspace{-.1cm}
+
\hspace{-.1cm}
\sup_{\theta\in\varTheta_1} \hspace{-.1cm}
\mathbb{P}_{\theta} \hspace{-.1cm}
\left( 
\max_{1\leq i\leq \N}
\left\lVert
\frac{1}{\T} \sum_{t=1}^\T 
(V_{it}^{11}
-
\mathbb{E}_\theta(
{V}_{it}^{11}))
\right\rVert_2
> 
\frac{\varepsilon}{2}
\right).
\end{split}
\end{equation*}		
To deal with the first of these terms let $\tilde{V}_{it}^{111}$ be the vector that collects all third-order derivatives with respect to $\varphi$ and let $\tilde{V}_{it}^{112}$ denote derivatives with respect to $\varphi$ (twice) and $\eta_i$. The tilde is used to indicate that these derivatives are evaluated at values $(\tilde{\varphi},\tilde{\eta}_i)$ that (elementwise) lie between $(\hat{\varphi},\hat{\eta}_i)$ and $({\varphi},{\eta}_i)$. A mean-value expansion around $(\varphi,\eta_i)$ yields
\begin{equation*}
\begin{split}
\left\lVert
\frac{1}{\T} \sum_{t=1}^\T 
(\hat{V}_{it}^{11}
-
{V}_{it}^{11})
\right\rVert_2
& \leq 
\frac{1}{\T} \sum_{t=1}^\T 
\left\lVert
\hat{V}_{it}^{11}
-
{V}_{it}^{11}
\right\rVert_2
\\
& \leq 
\frac{1}{\T} \sum_{t=1}^\T 
\left\lVert
\tilde{V}_{it}^{111}
\right\rVert_2
\,
\lVert
\hat{\varphi}
-
\varphi
\rVert_2
+
\frac{1}{\T} \sum_{t=1}^\T 
\left\lVert
\tilde{V}_{it}^{112}
\right\rVert_2
\,
\lVert
\hat{\eta}_i
-
\eta_i
\rVert_2
\\
& \leq 
\frac{1}{\T} \sum_{t=1}^\T 
\left\lVert
\tilde{V}_{it}^{111}
\right\rVert_1
\,
\lVert
\hat{\varphi}
-
\varphi
\rVert_2
+
\frac{1}{\T} \sum_{t=1}^\T 
\left\lVert
\tilde{V}_{it}^{112}
\right\rVert_1
\,
\lVert
\hat{\eta}_i
-
\eta_i
\rVert_2 .
\end{split}
\end{equation*}	
The uniform bound on the derivatives in Assumption \ref{ass:ass3}(ii) implies that
\begin{equation*}
\begin{split}	
\frac{1}{\T} \sum_{t=1}^\T 
\left\lVert
\tilde{V}_{it}^{111}
\right\rVert_1
& \leq 
(\dim \varphi)^3 \times \frac{1}{\T} \sum_{t=1}^\T b(z_{it})
,
\\
\frac{1}{\T} \sum_{t=1}^\T 
\left\lVert
\tilde{V}_{it}^{112}
\right\rVert_1
& \leq 
(\dim \varphi)^2 \times (\dim \eta_i) \times \frac{1}{\T} \sum_{t=1}^\T b(z_{it}).
\end{split}
\end{equation*}
Therefore, with $A\lesssim B$ meaning that there exists a finite constant $c$ such that $A \leq c\, B$, 
$$
\max_{1\leq i\leq \N}
\left\lVert
\frac{1}{\T} \sum_{t=1}^\T 
(\hat{V}_{it}^{11}
-
{V}_{it}^{11})
\right\rVert_2
\lesssim
\left(
\max_{1\leq i \leq \N}
\frac{1}{\T} \sum_{t=1}^\T b(z_{it})
\right)
\,
\left(\lVert \hat{\varphi}-\varphi \rVert_2 + \max_{1\leq i\leq \N} \lVert \hat{\eta}_i-\eta_i \rVert_2 \right).
$$
Now, the mixing conditions in Assumption \ref{ass:ass2} and the moment conditions on the bounding function $b$ in Assumption \ref{ass:ass3}(iii) imply that
$$
\sup_{\theta\in\varTheta_1}\mathbb{P}_{\theta}
\left(
\max_{1\leq i \leq \N}
\left\vert 
\frac{1}{\T} \sum_{t=1}^\T b(z_{it}) - \mathbb{E}_\theta(b(z_{it})) \right\rvert
>
\varepsilon
\right) = o(1)
$$
by an application of Lemma 1 of \cite{HahnKuersteiner2011} (which is easily extended to our setting; see the supplement). Also, 
$
\nicefrac{1}{\T} \sum_{t=1}^\T \mathbb{E}_\theta(b(z_{it}))
$
converges to its limit uniformly over $\varTheta_1$ by Assumption \ref{ass:ass3}(iv).
By Theorem 1 in \cite{KimSun2016}, 
$$
\sup_{\theta\in\varTheta_1}\mathbb{P}_{\theta}
\left(
\lVert \hat{\varphi}-\varphi \rVert_2
>
\varepsilon
\right) = o(1)
,
\qquad\sup_{\theta\in\varTheta_1}\mathbb{P}_{\theta}
\left(
\max_{1\leq i \leq \N}
\lVert \hat{\eta}_i-\eta_i \rVert_2
>
\varepsilon
\right) = o(1),
$$
and so 
$$
\sup_{\theta\in\varTheta_1}
\mathbb{P}_{\theta}
\left( 
\max_{1\leq i\leq \N}
\left\lVert
\frac{1}{\T} \sum_{t=1}^\T 
(\hat{V}_{it}^{11}
-
{V}_{it}^{11})
\right\rVert_2
> \frac{\varepsilon}{2}
\right)
 = o(1)
$$
follows. Next, again by Assumptions \ref{ass:ass2} and \ref{ass:ass3}, an application of (a uniform version of) Lemma 3 of \cite{HahnKuersteiner2011} (see the supplement) gives
$$
\sup_{\theta\in\varTheta_1}
\mathbb{P}_{\theta}
\left( 
\max_{1\leq i\leq \N}
\left\lVert
\frac{1}{\T} \sum_{t=1}^\T 
({V}_{it}^{11} 
-
\mathbb{E}_\theta({V}_{it}^{11}))
\right\rVert_2
> \frac{\varepsilon}{2}
\right)
 = o(1).
$$
Hence,
$$
\sup_{\theta\in\varTheta_1}
\mathbb{P}_{\theta}
\left( 
\max_{1\leq i\leq \N}
\left\lVert
\frac{1}{\T} \sum_{t=1}^\T 
(\hat{V}_{it}^{11}
-
\mathbb{E}_\theta(
{V}_{it}^{11}))
\right\rVert_2
> \varepsilon \,
\right)	= o(1),
$$
and the proof is complete. \qed

\setlength{\bibsep}{0pt} 
\bibliographystyle{chicago3} 
\bibliography{panel}

\end{document}